\pdfoutput=1
\documentclass[conference,letterpaper,10pt]{IEEEtran}

\usepackage[latin1]{inputenc}
\usepackage{times,amsmath}
\usepackage{amsfonts}
\usepackage{pstool}
\usepackage{subfigure}
\usepackage{multirow}
\usepackage{enumerate}
\usepackage{graphicx}
\usepackage{MnSymbol}
\usepackage{stfloats} 
\usepackage[table]{xcolor}
\usepackage[square, comma, sort&compress, numbers]{natbib}
\usepackage{nohyperref}
\usepackage{algorithm,algorithmic}
\usepackage{bigints}

\newcounter{tempEquationCounter} 
\newcounter{thisEquationNumber}

\makeatletter
\newcommand{\vast}{\bBigg@{4}}
\newcommand{\Vast}{\bBigg@{5}}
\makeatother

\begin{document}

\title{Channel Impulse Response-based Distributed Physical Layer Authentication}

\author{
\IEEEauthorblockN{ 
Ammar Mahmood\IEEEauthorrefmark{1}, Waqas Aman\IEEEauthorrefmark{1}, M. Ozair Iqbal\IEEEauthorrefmark{1}, M. Mahboob Ur Rahman\IEEEauthorrefmark{1}, Qammer H. Abbasi\IEEEauthorrefmark{2}  
}
\IEEEauthorblockA{
\IEEEauthorrefmark{1}Electrical engineering department, Information Technology University (ITU), Lahore, Pakistan \\\{ammar.mahmood,waqas.aman,ozair.iqbal,mahboob.rahman\}@itu.edu.pk 
}
\IEEEauthorblockA{
\IEEEauthorrefmark{2}Department of Electrical and Computer Engineering, Texas A\&M University at Qatar \\qammer.abbasi@qatar.tamu.edu
}

}

\maketitle

\begin{abstract} 

In this preliminary work, we study the problem of {\it distributed} authentication in wireless networks. Specifically, we consider a system where multiple Bob (sensor) nodes listen to a channel and report their {\it correlated} measurements to a Fusion Center (FC) which makes the ultimate authentication decision. For the feature-based authentication at the FC, channel impulse response has been utilized as the device fingerprint. Additionally, the {\it correlated} measurements by the Bob nodes allow us to invoke Compressed sensing to significantly reduce the reporting overhead to the FC. Numerical results show that: i) the detection performance of the FC is superior to that of a single Bob-node, ii) compressed sensing leads to at least $20\%$ overhead reduction on the reporting channel at the expense of a small ($<1$ dB) SNR margin to achieve the same detection performance. 

\end{abstract}

\section{Introduction}
\label{sec:intro}

Physical layer authentication addresses the problem of intrusion (impersonation) attack where a legitimate node Alice transmits its data to its intended recipient Bob, while Eve is an intruder node who unlawfully transmits on the channel allocated to Alice, in order to impersonate Alice before Bob. More precisely, for intrusion detection, Bob needs to authenticate each and every data packet it receives from the channel occupant (Alice or Eve). To carry out the (feature-based) authentication, Bob performs binary hypothesis testing while utilizing some device fingerprint (some physical layer attribute) as the decision metric. 

The (authentication) problem at hand is in principle very similar to the classical problems of detection, classification, clustering etc. Nevertheless, the only important distinction is that in case of authentication problem, the device fingerprint of Eve is unknown. This limitation renders the classical likelihood ratio test (LRT) not computable (and hence Bayesian methods are not directly applicable). Therefore, Neyman-Pearson based binary hypothesis testing is typically implemented to carry out the authentication task at Bob. 

As for the decision metric for binary hypothesis testing, Researchers have considered many different physical-layer attributes so far. Broadly speaking, these attributes can be classified into two categories: i) medium-based attributes, ii) hardware-based attributes. Medium-based attributes/fingerprints include different menifestations of the wireless channel such as channel frequency response \cite{Xiao:TWC:2008},\cite{Baracca:TWC:2012}, channel impulse response \cite{Wang:MILCOM:2011},\cite{Jitendra:COMSNETS:2010}, received signal strength \cite{Trappe:TPDS:2013}, angle-of-arrival \cite{Xiong:2010:SIGCOMM} etc. On the other hand, hardware-based attributes/fingerprints include carrier frequency offsets \cite{Wang:ICC:2012},\cite{Mahboob:Globecom:2014}, IQ-imbalance \cite{Wang:ICC:2014}, mismatch parameters of non-reciprocal hardware \cite{Mahboob:Arxiv:2016} etc. 

In this preliminary work, motivated by the classical distributed detection problem \cite{Varshney:IEEEProc:1997}, we study the {\it distributed authentication} problem. Specifically, we consider a situation where instead of a single Bob (sensor) node, there are $N$ Bob (sensor) nodes which all report their raw measurements or local decisions to a Fusion Center (FC) which makes the ultimate authentication decision. The need for Distributed Authentication arises whenever a single sensor (Bob) node alone can't be relied upon (e.g., it could come under deep fade/shadowing, it might have excessive AWGN/noise figure due to cheap receive circuitry etc.). Another potential application scenario is that of a Wireless Sensor Network \cite{Bauer:ACMTISS:2008} where the sensor (Bob) nodes are primitive in nature (i.e., they can't do sophisticated signal processing by themselves due to processor, storage and battery constraints); therefore, the sensor nodes simply relay the critical/sensitive data to the FC which carries out the ultimate authentication.

{\bf Contributions.} The main contributions of this paper are the following: i) a preliminary study on distributed authentication, ii) exploitation of compressed sensing at the Bob nodes for significant reduction of the reporting overhead to the FC. 

{\bf Outline.} The rest of this paper is organized as follows. Section-II introduces the system model. Section-III provides the necessary background for the channel impulse response which has been utilized as as device fingerprint in this work. Section-IV outlines the details of two strategies Bob (sensor) nodes could adopt to facilitate the cause of distributed detection. Section-V proposes to exploit compressed sensing at the Bob (sensor) nodes to reduce the overhead on the reporting channel. Section-VI provides some numerical results. Section VII concludes.

{\bf Notations.} Bold-face letters (e.g., $\mathbf{x}$) represent vectors; Capitalized bold-face letters (e.g., $\mathbf{X}$) represent matrices; $tr(.)$ denotes the trace operator; $(.)^T$ denotes the transpose operator; $(.)^H$ denotes the conjugate transpose operator; $\mathbb{E}(.)$ denotes the expectation operator; {\it i.i.d} implies independent and identically distributed; $\mathbf{0}$ represents a vector (of appropriate length) of all zero's; $blkdiag(\mathbf{X_1} \; \hdots \; \mathbf{X_N})$ represents a block diagonal matrix; $X \sim \mathcal{CN} (.)$ signifies that $X$ is a random variable with (circularly-symmetric) complex normal distribution; $||\mathbf{x}||_{l_p}$ represents the $l_p$-norm of vector $\mathbf{x}$.

\section{System Model}
\label{sec:sys-model}

\begin{figure}[ht]
\begin{center}
	\includegraphics[width=3in]{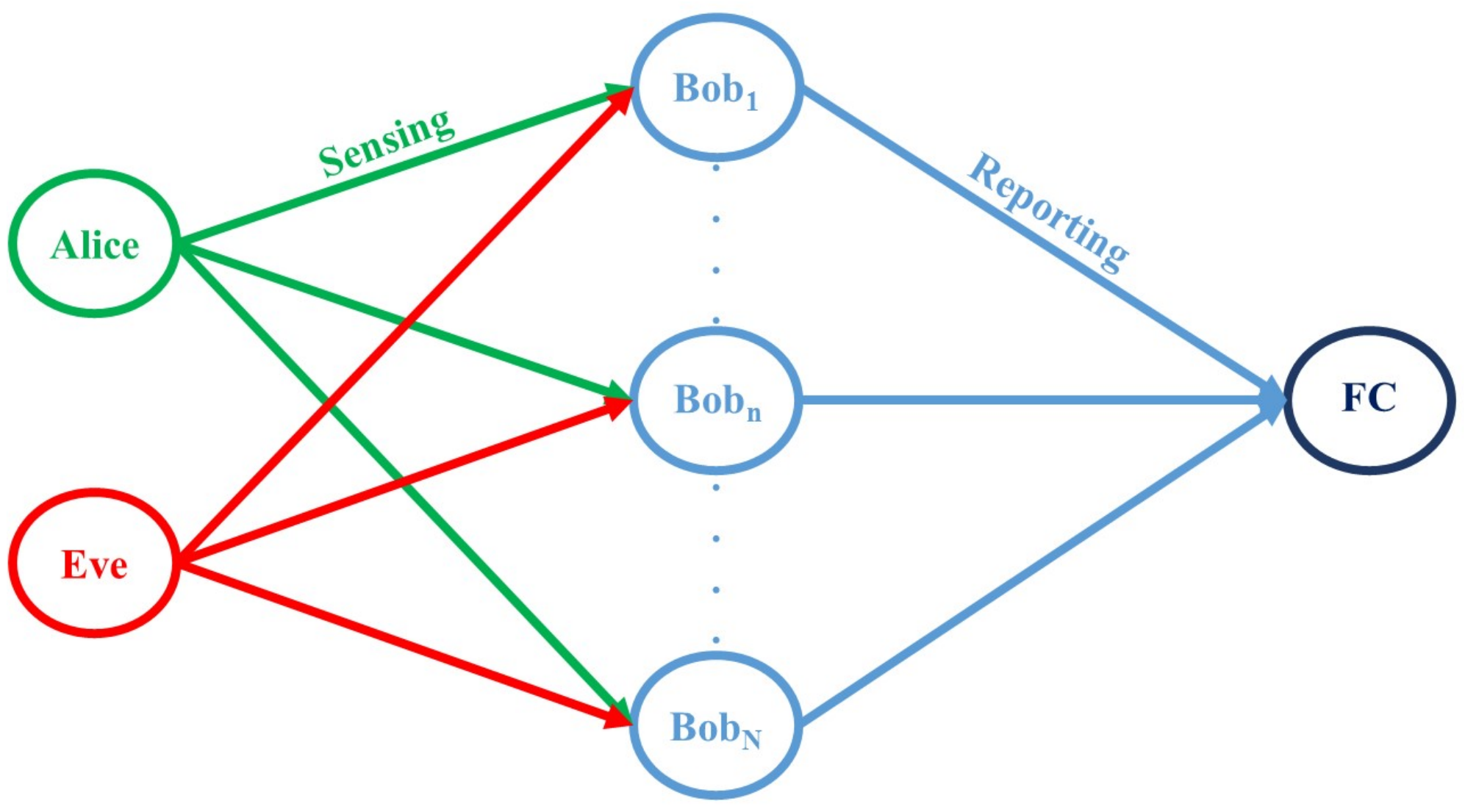} 
\caption{The Distributed Authentication problem.}
\label{fig:sysmodel}
\end{center}
\end{figure}

Following the spirits and motivation of distributed detection problem \cite{Varshney:IEEEProc:1997}, this paper studies the distributed authentication problem whereby $N$ number of Bob nodes simultaneously receive the packet sent by the channel occupant (either Alice or Eve) on the {\it sensing channel} (see Fig. \ref{fig:sysmodel}). The $N$ Bob nodes are either co-located (in the form of an antenna array), or, follow a random geometry (where each Bob node represents a different sensing device). In both scenarios, each Bob node reports some quantity (either its raw measurement, or, its local authentication decision) via a {\it reporting channel} to a Fusion Center (FC) which makes the ultimate authentication decision. Inline with previous literature \cite{Varshney:IEEEProc:1997}, this work assumes that the reporting channel is error-free, delay-free and time-slotted (the performance curves obtained under these assumptions simply become upper-bounds on the performance curves obtained under realistic settings where these assumptions don't hold). 

\section{Background: Channel Impulse Response as Device Fingerprint}
\label{sec:cir}

An intrusion (impersonation) attack occurs when an intruder node Eve unlawfully transmits on the (sensing) channel allocated to legitimate node Alice so as to impersonate Alice before Bob. To counter the challenge of intrusion attack, Bob needs to do authentication of each and every data packet it receives. Motivated by \cite{Jitendra:COMSNETS:2010}, this work utilizes the channel impulse response (CIR) as the device fingerprint for the purpose of authentication (at Bob nodes, or, Fusion center). More precisely, this work considers a single-carrier, wideband system with a time-slotted sensing channel (with $\tau$ seconds long timeslots). To introduce the notations, let's assume that Alice transmits on the shared sensing channel; then Bob $n$ sees the channel $\mathbf{h}_{A,B_n}^{(i.i.d)}$. More precisely, $\mathbf{h}_{A,B_n}^{(i.i.d)}$ is the channel impulse response, i.e., $\mathbf{h}_{A,B_n}^{(i.i.d)}=[h_{A,B_n}(1) \; h_{A,B_n}(2) \; \hdots \; h_{A,B_n}(L)]^T$ where $n=1,...,N$ and $L$ represents the total number of channel taps for the considered multipath (i.e., frequency-selective), time-invariant channel. Define:
\begin{equation*}
\mathbf{{H}_{AB}^{(i.i.d)}} 
\overset{\Delta}{=} [ \mathbf{h}_{A,B_1}^{(i.i.d)} \; \mathbf{h}_{A,B_2}^{(i.i.d)} \; \hdots \; \mathbf{h}_{A,B_N}^{(i.i.d)} ]   \in \mathbb{C}^{(L\times N)}    
\end{equation*}

Typically, each of the columns of $\mathbf{{H}_{AB}^{(i.i.d)}}$ is modelled as a Complex Gaussian vector; moreover, all the columns (channel vectors) are considered to be {\it i.i.d}. However, in this study, we consider a more general/realistic setting; i.e., we assume that the channels measured by all the $N$ Bob nodes are correlated (i.e., either all the Bob nodes are co-located, or, all the nodes in system model of Fig. \ref{fig:sysmodel} are located outdoors). Therefore, to make the columns of the matrix $\mathbf{{H}_{AB}^{(i.i.d)}}$ correlated with each other (to exploit the phenomena of compressed sensing later), we leverage the classical Kronecker-product based correlation model \cite{Kermoal:JSAC:2002}. For the system model in Fig. \ref{fig:sysmodel}, the Kronecker-product model reduces to the following:
\begin{equation}
\label{eq:Hcorr}
\mathbf{H_{AB}} = \frac{1}{\sqrt{tr(\mathbf{R_{B}})}} \mathbf{{H}_{AB}^{(i.i.d)}} \mathbf{R_{B}}^{1/2}
\end{equation}
where $\mathbf{R_{B}}$ $\in \mathbb{R}^{N\times N}$ represents the correlation matrix at the receive side (i.e., the correlation among Bob nodes). Specifically, we employ the exponential correlation model: $[\mathbf{R_{B}}]_{i,j}=\rho^{|i-j|}$ where $0\le \rho \le 1$ is the correlation amount and $i,j$ represent row and column indices of $\mathbf{R_{B}}$ (or, Bob $i$ and Bob $j$) respectively. Then, one can write: $\mathbf{{H}_{AB}} 
\overset{\Delta}{=} [ \mathbf{h}_{A,B_1} \; \mathbf{h}_{A,B_2} \; \hdots \; \mathbf{h}_{A,B_N} ]$.

Similarly, when Eve transmits on the shared sensing channel, then one can define:
\begin{equation*}
\mathbf{{H}_{EB}^{(i.i.d)}} 
\overset{\Delta}{=} [ \mathbf{h}_{E,B_1}^{(i.i.d)} \; \mathbf{h}_{E,B_2}^{(i.i.d)} \; \hdots \; \mathbf{h}_{E,B_N}^{(i.i.d)} ]  \in \mathbb{C}^{(L\times N)}     
\end{equation*}
The channel matrix $\mathbf{H_{EB}} \overset{\Delta}{=} [ \mathbf{h}_{E,B_1} \; \mathbf{h}_{E,B_2} \; \hdots \; \mathbf{h}_{E,B_N} ]$ with correlated columns is then constructed via Eq. (\ref{eq:Hcorr}) as well\footnote{At this point, it is worth mentioning that we haven't incorporated the transmit side correlation (between Alice's antenna and Eve's antenna) into the Kronecker product model. The reason for this is that, inline with the previous work \cite{Xiao:TWC:2008}, we assume that Alice and Eve are spaced far apart (more than one wavelength) so that their mutual correlation is negligible.}. 


\section{Distributed Authentication}
\label{sec:DistAuthen}

This section outlines the proposed distributed authentication framework which in turn utilizes the channel impulse response as the transmit device fingerprint.

During $m$-th timeslot, either Alice, or, Eve will transmit on the sensing channel (assuming that Eve avoids collisions so as to stay undetected). Therefore, when Bob $n$ receives a packet from the channel occupant (CO) (where CO $\in \{A,E\}$), it independently generates a measurement $\textbf{z}_n(m)$ of its local CIR $\mathbf{h}_{CO,B_n}$ as follows \cite{Jitendra:COMSNETS:2010}:
\begin{equation}
\label{eq:z_n}
\textbf{z}_n(m) = \mathbf{h}_{CO,B_n} + \textbf{v}_n(m)
\end{equation}
where $\textbf{v}_n(m)$ is the measurement noise. Specifically, $\textbf{v}_n(m) \sim \mathcal{CN}(\mathbf{0},\mathbf{\Sigma}_n)$ where $\mathbf{\Sigma}_n=\sigma_n^2 (\mathbf{S}^H\mathbf{S})^{-1}$ $\in \mathbb{R}^{L \times L}$; $\sigma_n^2$ denotes the variance/power of the AWGN at Bob $n$. Furthermore, we assume that $\textbf{v}_n(m)$ ($n=1,...,N$) are {\it i.i.d}\footnote{$\mathbf{S}$ is the matrix formed by the training symbols. See \cite{Jitendra:COMSNETS:2010} for more details.}.

With the raw measurements of local CIR available at each Bob node, two distinct strategies are possible: i) each Bob node sends its raw measurement as is (along with the ground truth) to the FC, ii) each Bob node does the authentication locally and sends its binary decision to the FC. Both strategies are discussed below.

\subsection{Each Bob node sends its raw measurement as-is to the Fusion Center}
Under this scenario, each Bob node sends its raw measurement $\mathbf{z}_{n}$ as-is (along with the ground truth) to the FC\footnote {For clarity of exposition, we drop the time-slot index $m$ in the rest of this paper.}. Since an error-free, time-slotted reporting channel is assumed, FC receives all the (perfect) measurements after $N$ time-slots (of the reporting channel). Fusion center then constructs a global measurement vector $\mathbf{{z}_{*}} \overset{\Delta}{=} [\mathbf{z}_{1}^T \; \mathbf{z}_{2}^T \; \hdots \; \mathbf{z}_{N}^T]^T$. Then, we can write: 
\begin{equation}
\label{eq:z*}
\mathbf{{z}_{*}}=\mathbf{h_{CO,B*}}+\mathbf{{v}_{*}}
\end{equation}
where $\mathbf{h_{CO,B*}}\overset{\Delta}{=}[\mathbf{h}_{CO,B_1}^T \; \mathbf{h}_{CO,B_2}^T \; \hdots \; \mathbf{h}_{CO,B_N}^T]^T$ and ${\mathbf{v_*}}\overset{\Delta}{=}[\mathbf{v}_1^T \; \mathbf{v}_2^T \; \hdots \; \mathbf{v}_N^T]^T$. ${\mathbf{v_*}}$ represents the global estimation error; ${\mathbf{v_*}}\sim \mathcal{CN} (\mathbf{0}, \mathbf{\Sigma_*})$ where $\mathbf{\Sigma_*} = blkdiag(\mathbf{\Sigma}_1 \; \mathbf{\Sigma}_2 \; \hdots \; \mathbf{\Sigma}_N)$ $\in \mathbb{R}^{(L\times N)\times (L\times N)}$. 

With $\mathbf{{z}_{*}}$ available, the FC casts the sender-node authentication problem as a binary hypothesis testing problem:

\begin{equation}
	\label{eq:H0H1_raw}
\begin{cases} 
H_0: & 
\mathbf{{z}_{*}}=\mathbf{h_{AB*}}+\mathbf{{v}_{*}}
\\
H_1: & 
\mathbf{{z}_{*}}=\mathbf{h_{EB*}}+\mathbf{{v}_{*}}
\end{cases}
\end{equation}
where ${\mathbf{h_{AB*}}}=[\mathbf{h}_{A,B_1}^T ,\mathbf{h}_{A,B_2}^T ,...,\mathbf{h}_{A,B_N}^T ]^T$ and ${\mathbf{h_{EB*}}}=[\mathbf{h}_{E,B_1}^T ,\mathbf{h}_{E,B_2}^T ,...,\mathbf{h}_{E,B_N}^T ]^T$. Then, ${\mathbf{z_*}|H_0} \sim \mathcal{CN}({\mathbf{h_{AB*}}}, \mathbf{\Sigma_*})$ and ${\mathbf{z_*}|H_1} \sim \mathcal{CN}({\mathbf{h_{EB*}}}, \mathbf{\Sigma_*})$. If $H_0=1$, received data (on the sensing channel) is accepted by the FC; if $H_1=1$, received data is discarded by the FC.

Next, assuming that $\mathbf{h_{AB*}}$ is known to the FC (via the prior training of the Bob nodes, on a secure channel) with sufficient accuracy, the FC applies the following test \cite{Jitendra:COMSNETS:2010}:
\begin{equation}
	\label{eq:test_raw}
	 (\textbf{z}_* - \mathbf{h_{AB*}})^H \mathbf{\Sigma_*}^{-1} (\textbf{z}_* - \mathbf{h_{AB*}}) \underset{H_0}{\overset{H_1}{\gtrless}} \delta  
\end{equation}
where $\delta$ is the comparison threshold whose value is to be determined by the FC. This work follows Neyman-Pearson procedure to systematically compute the threshold $\delta$. Specifically, $\delta$ is computed from a pre-specified, maximum Type-1 error (i.e., probability of false alarm, $P_{fa}$) that the FC can tolerate. Then, for a given Type-1 error, Neyman-Pearson method guarantees to minimize the Type-2 error (i.e., probability of missed detection, $P_{md}$).   

Denote by $T$ the test statistic in Eq. (\ref{eq:test_raw}); i.e., $T=(\textbf{z}_* - \mathbf{h_{AB*}})^H \mathbf{\Sigma_*}^{-1} (\textbf{z}_* - \mathbf{h_{AB*}})$. Then, the test statistic $T|H_0 \sim \chi^2(2NL)$; i.e., $T$ has central Chi-squared distribution with $2NL$ degrees of freedom, under $H_0$. Then, the probability of false alarm $P_{fa}$ (i.e., incorrectly identifying Alice's packet as if it is from Eve) is given as:
\begin{equation} \label{eq:pfa_fc}
	P_{fa} = Pr(T>\delta |H_0) = \int_{\delta}^\infty p_{T|H_0}(x) dx
\end{equation}
where $p_{T|H_0}(x) \sim \chi^2(2NL)$ is the probability density function of $T|H_0$. Thus, one can set $P_{fa}$ to some value $\alpha$ in Eq. (\ref{eq:pfa_fc}) and solve for the threshold $\delta$.

With $P_{fa}=\alpha$, the performance of the hypothesis test in Eq. (\ref{eq:test_raw}) is solely characterized by the Type-2 error a.k.a the probability of missed detection: $P_{md}=Pr(T<\delta |H_1)$. Since computing the distribution of $T|H_1$ is quite involved, we numerically compute the value of $P_{md}$ in simulation section.

\subsection{Each Bob node sends its local decision to the Fusion Center}
Under this scenario, Bob node $n$ (independently) utilizes its measurement $\textbf{z}_n$ of its local CIR $\mathbf{h}_{CO,B_n}$ (see Eq. (\ref{eq:z_n})) to construct the following binary hypothesis testing problem:
\begin{equation}
	\label{eq:H0H1_local_D}
	 \begin{cases} H_0: & \textbf{z}_n = \mathbf{h}_{A,B_n} + \textbf{v}_n \\ 
                                 H_1: & \textbf{z}_n = \mathbf{h}_{E,B_n} + \textbf{v}_n \end{cases}
\end{equation}

Once again, assuming that $\mathbf{h}_{A,B_n}$ is perfectly known to Bob $n$, it applies the following test \cite{Jitendra:COMSNETS:2010}:
\begin{equation}
	\label{eq:test_LD}
	 (\textbf{z}_n - \mathbf{h}_{A,B_n})^H \mathbf{\Sigma}_n^{-1} (\textbf{z}_n - \mathbf{h}_{A,B_n}) \underset{H_0}{\overset{H_1}{\gtrless}} \delta_n   
\end{equation}
where $\delta_n$ is the comparison threshold whose value is to be determined by Bob $n$. Once again, we follow the Neyman-Pearson procedure to systematically compute the threshold $\delta_n$.   

Denote by $T_n$ the test statistic in Eq. (\ref{eq:test_LD}); i.e., $T_n=(\textbf{z}_n - \mathbf{h}_{A,B_n})^H \Sigma_n^{-1} (\textbf{z}_n - \mathbf{h}_{A,B_n})$. Then, the test statistic $T_n|H_0 \sim \chi^2(2L)$; i.e., $T_n$ has central Chi-squared distribution with $2L$ degrees of freedom, under $H_0$. Then, the probability of false alarm $P_{fa,n}$ is given as:
\begin{equation} \label{eq:pfa_1}
	P_{fa,n} = Pr(T_n>\delta_n |H_0) = \int_{\delta_n}^\infty p_{T_n|H_0}(x) dx
\end{equation}
where $p_{T_n|H_0}(x) \sim \chi^2(2L)$ is the probability density function of $T_n|H_0$. Thus, one can set $P_{fa,n}$ to some value $\alpha_n$ in Eq. (\ref{eq:pfa_1}) and solve for the threshold $\delta_n$.

With $P_{fa,n}=\alpha_n$, the performance of the hypothesis test in Eq. (\ref{eq:test_LD}) is solely characterized by the Type-2 error a.k.a the probability of missed detection: $P_{md,n}=Pr(T_n<\delta_n |H_1)$. Since it is difficult to compute the distribution of $T_n|H_1$, we numerically compute the value of $P_{md,n}$ in simulation section.

Then, the Fusion Center, having received the vector of local (binary-valued, hard) decisions $\mathbf{u_*}$ by all the Bob nodes, applies the following (heuristic) fusion rules to generate the ultimate (binary-valued, hard) decision: i) OR (optimistic) rule, ii) AND (pessimistic) rule, iii) majority voting, iv) weighted averaging.

\section{Overhead Reduction on Reporting Channel via Compressed Sensing}
\label{sec:CS}

This section attempts to invoke Compressed Sensing\footnote{The interested reader is referred to \cite{Candes:SPMag:2008} for a comprehensive overview of compressed sensing.} at the Bob nodes owing to the fact that the CIR measurements made by the Bob nodes are correlated (see Eq. \ref{eq:Hcorr}). To this end, this section focuses on a specialized system model (more than the one given in Fig. \ref{fig:sysmodel} earlier) whereby the $N$ ($N$ is large) Bob nodes are all co-located as a linear antenna-array on a single receive device, now-called the relay node. Such system model could represent, for example, a situation where a multi-antenna relay node (e.g., a massive MIMO base station) forwards the data of a source (with unknown identity) towards a destination (the FC). Moreover, the relay node can either stay indifferent/blind to the authentication cause of the FC (the raw measurements case), or, the relay node could help the authentication cause of the FC (the local decisions case). We further assume that the relay node doesn't know its (vector) channel towards the FC. Thus, the multi-antenna relay doesn't do beamforming towards the FC; it rather transmits the length-$NL$ vector (containing raw measurements or local decisions) to the FC in a sequential manner. All in all, this specialized system model allows us to invoke compressed sensing (CS) at the relay node (to compress the length-$NL$ report vector into a length-$M$ vector where $M<<NL$) to reduce the number of channel uses (from $N$ to $M$) on the reporting channel.

At this point, it is worth mentioning that to exploit compressed sensing under the generalized system model shown in Fig. \ref{fig:sysmodel}, one needs to employ more sophisticated medium access schemes on the reporting channel (which is beyond the scope of the current work)\footnote{For example, \cite{Yang:TSP:2013} presents a scheme where during each of the $M$ time-slots, each of the $N$ sensor nodes transmits its local measurement to the FC with some probability $p$, thus making the sensing/measurement matrix $\mathbf{\Phi}$ non-Gaussian.}.

We now recall from the previous section that having received the data from the (sensing) channel occupant (Alice or Eve), the relay node could adapt either of the two distinct {\it compress-and-forward} strategies: i) the relay node compresses the vector of raw measurements and sends it to the FC (along with the vector of ground truths), ii) the relay node compresses the vector of local decisions and sends it to the FC. Both strategies are discussed below.

\subsection{The relay node compresses the vector of raw measurements and sends it to the Fusion Center}
The relay node compresses the vector of raw measurements $\mathbf{z_*}$ $\in \mathbb{C}^{(NL\times 1)}$ (see Eq. \ref{eq:z*}) into a vector $\mathbf{y}$ $\in \mathbb{C}^{(M\times 1)}$ ($M<<NL$) via random projections method:
\begin{equation}
\mathbf{y} = \mathbf{\Phi}\mathbf{z_*}
\end{equation}
where $\mathbf{\Phi}$ $\in \mathbb{R}^{(M\times NL)}$ is a random matrix, the so-called sensing/measurement matrix, whose entries are typically the realizations of {\it i.i.d} Gaussian or Bernoulli variables. 

After the compression, the relay node transmits the vector $\mathbf{y}$ to the FC in $M$ time-slots. Due to the error-free reporting channel, the FC receives $\mathbf{y}$ as-is; therefore, its task becomes to recover an estimate $\hat{\mathbf{z}}_*$ of $\mathbf{z_*}$ from $\mathbf{y}$. To this end, we invoke the classical Orthogonal Matching Pursuit (OMP) algorithm at the FC \cite{Tropp:TIT:2007}. Specifically, the OMP solves the following $l_1$-norm minimization problem:
\begin{equation}
\label{eq:omp}
\min ||\mathbf{z_0}||_{l_1} \; s.t.  \; \mathbf{y} = \mathbf{\Phi} \mathbf{\Psi}^T \mathbf{z_0}
\end{equation}
where 
\begin{equation}
\label{eq:sparse}
\mathbf{z_0} = \mathbf{\Psi} \mathbf{z}_*
\end{equation} 
is the actual $K$-sparse representation of $\mathbf{z_*}$ in a transform domain. The matrix $\mathbf{\Psi}$ is called the sparsifying basis.
The most frequently used sparsifying basis $\mathbf{\Psi}$ include discrete fourier transform (DFT), discrete cosine transform (DCT), Karhunen-Loeve Transform (KLT) etc. Finally, when $\hat{\mathbf{z}}_*$ is available via Eqs. (\ref{eq:omp}),(\ref{eq:sparse}), the FC follows the procedure of Section-IV-A to carry out the authentication.

At this point, it is necessary to emphasize that the invocation of compressing sensing at the relay node is feasible only when the vector $\mathbf{z_*}$ is $K$-sparse (for some $K<<NL$) in some transform domain. This condition is fulfilled when the channel under consideration exhibits spatial correlation (see Eq. \ref{eq:Hcorr}) \cite{Ping:WCNC:2012}. Typically, the reconstruction error $\mathcal{E}=\mathbb{E}[||\hat{\mathbf{z}}_* - \mathbf{z_*}||_{l_2}^2]$ decreases with increase in $\rho$ (see Eq. \ref{eq:Hcorr}) and vice versa.

\subsection{The relay node compresses the vector of local decisions and sends it to the Fusion Center}
Under this strategy, the relay node compresses the vector of local decisions $\mathbf{u_*}$ $\in \{0,1\}^{(NL\times 1)}$ into a vector $\mathbf{y}$ $\in \{0,1\}^{(M\times 1)}$ ($M<<NL$) via random projections method and transmits $\mathbf{y}$ to the FC in $M$ time-slots, as discussed above. The FC then recovers $\hat{\mathbf{u}}_*$ from $\mathbf{y}$ via OMP and follows the procedure of Section-IV-B to carry out the authentication task.
\section{Numerical Results}
\label{sec:results}

We assume uniform power delay profile (PDP) to generate each of the $N$ CIRs (with $L=6$ taps each) on the sensing channel. The received SNR at Bob $n$ is defined as $1/\sigma_n^2$ where $\sigma_n^2$ is the power of AWGN at Bob $n$. Furthermore, we evaluate the case of homogeneous SNRs only (i.e., we assume that the SNR/link quality of all the $N$ sensing links is the same). Finally, in all the results below, we evaluate the detection probability $P_d$ (where $P_d=1-P_{md}$) of the FC as the SNR (of a single sensing link) is varied over its full operational range. 

\begin{figure}[ht]
\begin{center}
	\includegraphics[width=3.5in]{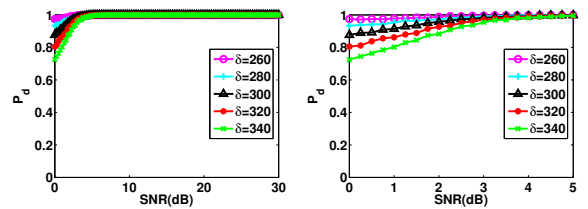} 
\caption{Detection performance of the FC when Bob nodes send their Raw Measurements to FC (right-side plot shows the zoomed-in view (corresponding to SNR $\in$ $0-5$ dB) of the left-side plot).}
\label{fig:DA-RM}
\end{center}
\end{figure}

Fig. \ref{fig:DA-RM} represents the case when Bob nodes send their Raw Measurements to the FC. For Fig. \ref{fig:DA-RM}, we set $N=10$; the threshold $\delta$ is varied over the range: $260:20:340$. Fig. \ref{fig:DA-RM} shows that with decrease in $\delta$ (or, equivalently, with increase in the false alarm rate $P_{fa}$), the detection probability increases as one would intuitively expect. Additionally, the detection performance remains stable to a high value (close to 1) for SNR (at the FC) as low as $5$ dB. 

\begin{figure}[ht]
\begin{center}
	\includegraphics[width=4in]{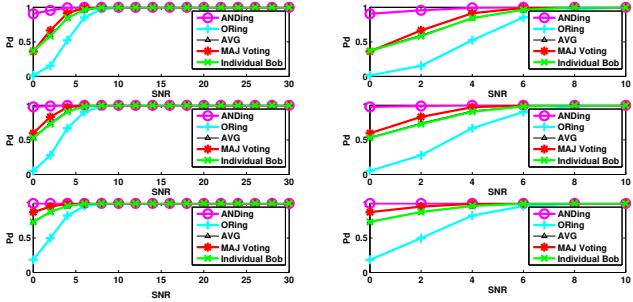} 
\caption{Bob nodes send their Local Decisions to the FC (right-side plot shows the zoomed-in view (corresponding to SNR $\in$ $0-10$ dB) of the left-side plot).}
\label{fig:DA-LD}
\end{center}
\end{figure}

Fig. \ref{fig:DA-LD} represents the case when Bob nodes send their Local Decisions to FC. In Fig. \ref{fig:DA-LD}, $N=10$; additionally, from top to bottom, each of the $N$ thresholds is varied as: $\delta_n$ $\in$ $\{39,32.9,26.2\}$ which maps to the false alarm rates: $P_{fa,n}$ $\in$ $\{0.0001,0.001,0.01\}$ respectively. Fig. \ref{fig:DA-LD} shows that the AND-ing (OR-ing) based pessimistic (optimistic) fusion rule performs the best (worst) while the performance of the Majority Voting (Averaging) fusion rule is slightly better than (same as) the performance achieved by a single Bob node. Additionally, from top to bottom, sacrifice in terms of more and more false alarm rate results in improvement of detection performance of all the schemes for any given SNR.

\begin{figure}[ht]
\begin{center}
	\includegraphics[width=4in]{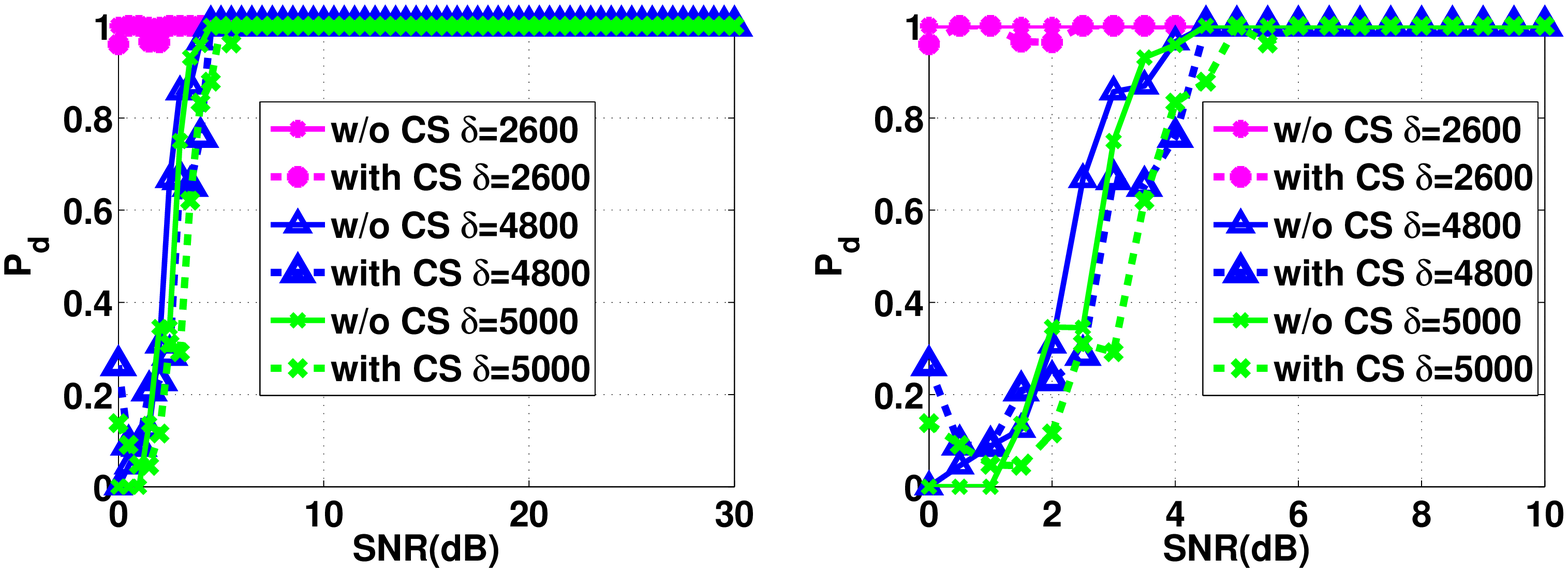} 
\caption{Bob nodes send their raw measurements to the FC via compressed sensing (right-side plot shows the zoomed-in view (corresponding to SNR $\in$ $0-10$ dB) of the left-side plot).}
\label{fig:DA-RM-CS}
\end{center}
\end{figure}

\begin{figure}[ht]
\begin{center}
	\includegraphics[width=4in]{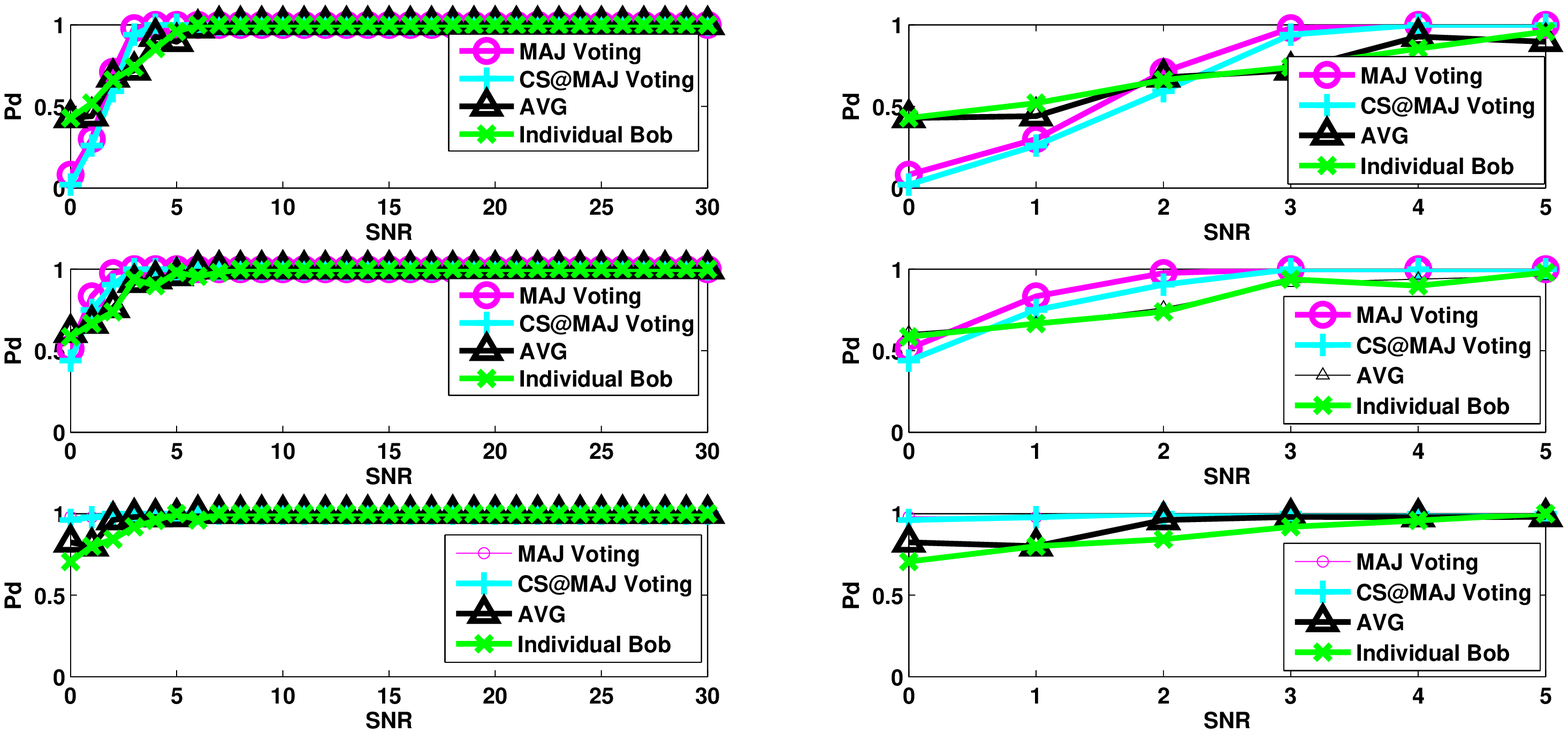} 
\caption{Bob nodes send their local decisions to the FC via compressed sensing (right-side plot shows the zoomed-in view (corresponding to SNR $\in$ $0-5$ dB) of the left-side plot).}
\label{fig:DA-LD-CS}
\end{center}
\end{figure}

Figs. \ref{fig:DA-RM-CS}, \ref{fig:DA-LD-CS} show the results equivalent to Figs \ref{fig:DA-RM}, \ref{fig:DA-LD} when the Bob nodes (a.k.a the relay node) invoke the compressed sensing before sending their measurements to the FC. For compressed sensing, we have used a Gaussian measurement matrix $\mathbf{\Phi}$ and DCT as the sparsifying basis $\mathbf{\Psi}$. 

Fig. \ref{fig:DA-RM-CS} represents the case when Bob nodes send their raw measurements to the FC after compressed sensing. In Fig. \ref{fig:DA-RM-CS}, we set $N=100$; the threshold $\delta$ is varied in the range: $\delta$ $\in$ $\{2600,4800,5000\}$. The compressed sensing compresses the length $N\times L=100\times 6=600$ measurement vector $\mathbf{z_*}$ into a length $M=480$ vector achieving a saving of $20\%$ in reporting overhead. Fig. \ref{fig:DA-RM-CS} basically compares the two strategies by the relay node: S1) the relay node does the CS prior to sending the raw measurement vector to the FC, S2) the relay node sends the raw measurement vector as-is to the FC. Fig. \ref{fig:DA-RM-CS} indicates that a small ($<1$ dB) SNR margin is needed by S1 to achieve the same detection performance as S2, due to finite reconstruction error $\mathcal{E}$ at the FC.  

Fig. \ref{fig:DA-LD-CS} represents the case when Bob nodes send their local decisions to the FC after compressed sensing. In Fig. \ref{fig:DA-LD-CS}, we set $N=100$; additionally, from top to bottom, each of the $N$ thresholds $\delta_n$ is varied in the range: $\delta_n$ $\in$ $\{39,32.9,26.2\}$ which maps to the following false alarm rates: $P_{fa,n}$ $\in$ $\{0.0001,0.001,0.01\}$ respectively. In all the three plots, the performance of the majority voting based fusion rule is considered with and without compressed sensing at the Bob nodes. In this case, compressed sensing at the Bob nodes leads to a $30\%$ overhead reduction on the reporting channel. 

\section{Conclusion}
\label{sec:conclusion}

We did a preliminary study on the problem of distributed authentication in wireless networks. Specifically, we considered a system where multiple Bob (sensor) nodes listen to a channel and report their {\it correlated} measurements to a Fusion Center (FC). Numerical results showed that the authentication performance of the FC is superior to that of a single Bob node. Additionally, the correlated measurements by the Bob nodes allowed us to invoke compressed sensing at the Bob nodes to reduce the reporting overhead to the FC by at least $20\%$. Immediate future work will investigate the design of: i) optimal decision rules at the Bob nodes and the FC, ii) clever medium access schemes on the reporting channel.

\appendices

\section*{Acknowledgements}
This publication was made possible by NPRP grant $\# 7-125-2-061$ from the Qatar National Research Fund (a member of Qatar Foundation). The statements made herein are solely the responsibility of the authors.

\footnotesize{
\bibliographystyle{IEEEtran}
\bibliography{references}
}

\vfill\break

\end{document}